\documentclass{PoS}
\usepackage{subfigure}
\newcommand{\wh}{\widehat}
\def\Im{{\rm Im} }

\def\beq {\begin{equation}}
\def\eeq {\end{equation}}
\newcommand{\as}{\alpha_s}

\title{On the perturbative expansion of $\tau$ hadronic spectral
 function moments}

\ShortTitle{Perturbative expansion of $\tau$ hadronic spectral
 function moments}

\author{\speaker{Diogo Boito}\footnote{TUM-HEP-873/13}\\
        Physik Department T31, Technische Universit\"at M\"unchen, \\
James-Franck-Stra\ss e 1, D-85748 Garching, Germany\\
        E-mail: \email{diogo.boito@tum.de}}

\abstract{In the determination of $\alpha_s$ from tau decays several different moments of the hadronic
spectral functions have been used. In a recent work, we performed an analysis of their perturbative 
behaviour under two different assumptions for the higher-order coefficients of the Adler function. We
showed that the various moments can be divided in a small number of classes. We concluded
that some of the moments commonly employed in $\alpha_s$ extractions should be avoided due to their bad 
perturbative behaviour. Furthermore, for the moments that have a good perturbative behaviour, and under 
reasonable assumptions for the higher-order behaviour of the Adler function, fixed-order perturbation 
theory (FOPT) provides the superior framework for the renormalization group  improvement.
Here we discuss an extension of this analysis where we consider the perturbative series
for values of the hadronic invariant mass squared $s_0\leq m_\tau^2$. Our conclusions are not altered
within a reasonable $s_0$ window.}

\FullConference{Xth Quark Confinement and the Hadron Spectrum,\\
		October 8-12, 2012\\
		TUM Campus Garching, Munich, Germany}

\begin{document}

\section{Introduction}

In the last 20 years, hadronic tau decays have been an important source of empirical information
on fundamental parameters of QCD. Notably, the strong coupling, $\alpha_s$, can be extracted
with a good precision at relatively low energies, close to the edge of the validity of perturbative QCD.
After the measurement of the spectral functions  at LEP, other parameters such as the 
strange quark mass, the CKM matrix element $V_{us}$, as well as non-perturbative condensates 
could be extracted (see e.g.~\cite{DHZ05}).
The extraction of these parameters relies on sum rules.
Quark-hadron duality and the optical theorem allow to express the decay rate as weighted integrals of
the vector and axial-vector spectral functions running over the hadronic invariant mass squared $s$ from 
threshold up to  $m_\tau^ 2$.   

Using the analytic properties of the quark-antiquark correlators,  the theoretical counter-part of the experimental  quantities are expressed as contour-integrals in the complex energy plane with fixed~$|s|=m_\tau^ 2$.  However, in the theoretical description of $\tau$ decays two main  obstacles remain. The first  is referred to as 
{\it duality violations} (DVs). They are related to the break-down of local quark-hadron duality in the vicinity of 
the Minkowski axis (where resonance effects 
become important). In the past, they have been neglected
due to a fortuitous kinematical suppression  of the problematic region in the contour integration. Recently, thanks 
to the progress in modelling DVs realistically~\cite{DVsRussians,CGP}, they have been included self-consistently in 
a full-fledged $\alpha_s$  analysis~\cite{AlphasDVs2011,AlphasDVs2012}. The second important obstacle is the 
prescription for the renormalization group (RG) improvement of the perturbative series. The most widely employed 
prescriptions are fixed-order perturbation theory (FOPT)~\cite{Jamin05,BJ2008} and contour improved perturbation 
theory (CIPT)~ \cite{CIPT,PLD92}. When used in practice, they lead to 
different $\alpha_s$ results. With the recently computed $\alpha_s^4$ 
correction~\cite{bck08}, the difference  became even more pronounced. Several works have dealt with this 
discrepancy~\cite{BJ2008,SM,CF,DGM,Cvetic:2010ut,AAC} in the light of the $\alpha_s^ 4$ term. The conclusions in favour of FOPT or CIPT (or a third prescription) are based on (implicit or explicit) assumptions  on the yet unknown higher order $\alpha_s$ corrections. In this context,
the goal of Ref.~\cite{BJ2008} was to construct a plausible  model for the higher-order corrections of the Adler
function from the leading renormalon singularities of its Borel transform, using only general RG arguments 
to describe the structure of the singularities in the Borel plane. After matching the model to the known 
coefficients in QCD, the main conclusion of Ref.~\cite{BJ2008} was that FOPT is to be preferred over CIPT, since 
FOPT provides a closer approach to the Borel resummed results --- in the spirit of an asymptotic series.

This conclusion  was based solely on the analysis of the weight $w_\tau(x)$, obtained from the kinematics of the decay. This  is not entirely satisfactory since realistic determinations of $\alpha_s$
employ (and often require) several different weight functions $w_i(x)$. In fact, any analytical $w_i(x)$ gives
rise to a valid sum-rule that emphasises a given part of the spectral functions, as well as different
contributions in the theoretical description. In the literature, several weight-functions have been employed and 
yet little attention has been paid to the moment dependence of the convergence properties of the perturbative 
series. We have addressed this question in Ref.~\cite{BBJ2012} and pursued the 
FOPT/CIPT comparison for several weight functions. We showed that the different moments employed in the literature 
can be divided in a small number of categories. The characteristics of their perturbative series 
could be linked to generic features of the moment weight function and the dominant 
renormalon singularities of the Adler  function. We concluded that some of the moments currently employed 
in some  $\alpha_s$ extractions should be avoided due to the poor convergence of their perturbative expansions. Additionally, for all moments that display good perturbative behaviour --- and under reasonable assumptions for the higher-order corrections --- FOPT provided the best framework to the RG improvement. 

A point that was not discussed in~\cite{BBJ2012} is the stability of the conclusions with 
respect to  $s_0$ variations ($s_0$ being the upper limit 
of integration in the sum-rule, see Eq.~(\ref{Rtauth})). The relevance of this issue lies in the fact 
that several $\alpha_s$ analyses use sum-rules where the data are integrated up to $s_0<m_\tau^2$. Here we show that the conclusions of~\cite{BBJ2012} remain valid  when $s_0$ is varied away from $m_\tau^ 
2$.

\section{Theoretical framework, model, and results}

We work with generalized sum-rules, where the weight function in the integrals can be any analytical 
function $w_i(x)$ and the upper limit of integration is taken to be any point $s_0\leq m_\tau^2$.
The experimental side of the sum-rules are then written as integrals over the spectral functions as
\begin{equation}
\label{Rtauth}
R_{\tau,{V/A}}^{w_i}(s_0) \,=\, 12\pi S_{\rm EW} |V_{ud}|^2 \!\int_0^{s_0}
\frac{ds}{s_0}\,\biggl( 1-\frac{s}{s_0}\biggr)^{\!2}
\biggl(1+2\frac{s}{s_0}\biggr)\biggl[
\Im\,\Pi^{(1+0)}_{V/A}(s)-\frac{2s}{s_0+2s}\Im\,\Pi^{(0)}_{V/A} (s) \,\biggr] \,.
\end{equation}
The two point functions are defined as $\Pi_{V/A}^{\mu\nu}(p) \,\equiv\,  i\!\int \! dx \, e^{ipx} \,
\langle\Omega|\,T\{ J_{V/A}^{\mu}(x)\,J_{V/A}^{\nu}(0)^\dagger\}|
\Omega\rangle$ and they assume the usual decomposition into longitudinal and transversal
components. The  $V$ and $A$ currents are
given by $J_{V/A}^{\mu}(x)= (\bar u\gamma^\mu(\gamma_5) d)(x)$. 

The theoretical counter-part of Eq.~(\ref{Rtauth}) is obtained from the counter-clock wise contour integration
of the correlators with $|s|=s_0$. The contributions on the theory side can 
be organized~as 
\begin{equation}
\label{RtauDeltas}
R_{V/A}^{w_i}(s_0) \,=\, \frac{N_c}{2}\,S_{EW} |V_{ud}|^2 \biggl[\,
\delta^{\rm tree}_{w_i} + \delta^{(0)}_{w_i}(s_0) + 
\sum_{D\geq 2}\delta^{(D)}_{w_i,V/A}(s_0) +\delta^{\rm DV}_{w_i,V/A}(s_0)
\,\biggr] \,.
\end{equation}
The perturbative contribution is contained in $\delta^{\rm tree}_{w_i}$ and $\delta^{(0)}_{w_i}$, of which 
$\delta^{(0)}_{w_i}$ contains the loop corrections.  In the chiral limit they
are the same for $V$ and $A$ correlators, and correspond to the
perturbative series of $\Pi^{(1+0)}_{V/A}(s)$. The quark-mass corrections,
as well as contributions from operators with $D>2$ in the OPE, are encoded 
in the terms $\delta^{(D)}_{w_i,V/A}$;  DV contributions
are represented by $\delta^{\rm DV}_{w_i,V/A}$.

Our focus is on the behaviour of the perturbative correction and it is convenient
to write it in terms of the RG invariant Adler function, whose expansion in 
$\alpha_s$ can be written as
\begin{equation}
\label{D10D0}
D^{(1+0)}(s) \,\equiv\, -\,s\,\frac{d}{ds}\,\Pi^{(1+0)}(s) =\, \frac{N_c}{12\pi^2} \sum\limits_{n=0}^\infty
a_\mu^n \sum\limits_{k=1}^{n+1} k\, c_{n,k}\,\left( \log \frac{-s}{\mu^2}\right)^{k-1},
\end{equation}
where $a_\mu=\alpha(\mu)/\pi$. RG invariance implies that only the coefficients $c_{n,1}$ are independent. The other $c_{n,k}$ can be expressed in terms of  $c_{n,1}$ and $\beta$-function coefficients. The perturbative contribution to the theory side of the sum-rules is then 
\begin{equation}
\label{del0}
\delta^{(0)}_{w_i} \,=\, \sum\limits_{n=1}^\infty  \sum\limits_{k=1}^{n}
k\,c_{n,k} \;\frac{1}{2\pi i}\!\!\oint\limits_{|x|=1} \!\! \frac{dx}{x}\,
W_i(x) \log^{k-1}\biggl(\frac{-s_0 x}{\mu^2}\biggr)a_\mu^n \,,
\end{equation}
with $x=s/s_0$ and $W_i(x) = 2\int_x^ 1 dz\, w_i(z)$. Due to the RG invariance of $D^{(1+0)}(s)$ one has the freedom of setting
the scale $\mu$. The FOPT prescription corresponds to the choice $\mu^2=s_0$. In 
this case, the coupling $a(s_0)$ is taken out-side the integrals and one is left 
with the integration of powers of $\log(-x)$. The CIPT choice correspond to $\mu^ 2=-s_0x$,
which resums the logarithms but, in turn, the integrals are  done (numerically)
over the running coupling $a(-s_0x)$. 

In order to compare FOPT and CIPT as well as understand the perturbative behaviour of spectral
function moments, one must have an ansatz for the unknown higher-order Adler function coefficients $c_{n,1}$.
Here we follow the method introduced in Ref.~\cite{BJ2008} which makes use of the available knowledge of the 
renormalon structure of the Borel transformed Adler function. The idea is to construct
a realistic model for the Borel transform using the leading singularities.  We work with the function $\wh D(s)$
and its Borel transform, $B[\wh D](t)$, defined as
\beq
\frac{12\pi^2}{N_c}\,D^{(1+0)}_V(s) \,\equiv\, 1 + \wh D(s) \,\equiv\, 1 +
\sum\limits_{n=0}^\infty r_n \,\as(\sqrt{s})^{n+1} \,, \qquad B[\wh D](t) \,\equiv\, \sum\limits_{n=0}^\infty r_n\,\frac{t^n}{n!}.
\eeq
The original series can be understood as an asymptotic expansion of the inverse of $B[\wh D](t)$,
\beq
 \wh D(\alpha) \,\equiv\, \int_0^\infty dt\,{\rm e}^{-t/\alpha}\,
B[\wh D](t)\,,\label{BorelSum}
\eeq
when the integral exists. Singularities of $B[\wh D](t)$ on the positive real axis (infra red (IR) 
renormalons) give rise to fixed-sign asymptotic series and obstruct the Borel summation, Eq.~(\ref{BorelSum}). This 
introduces an ambiguity in the integral that is expected to be cancelled against exponentially small terms 
in $\alpha_s$, or power corrections(due to the logarithmic running of the coupling). Singularities on the 
negative real axis (ultra violet renormalons (UV)) give rise to  sign-alternating series.


General RG arguments and the structure of the OPE allow one to determine the position and strength of the 
renormalon singularities in the $t$ plane, though not their residues~\cite{Renormalons}.  The fixed-sign
nature of the exactly known coefficients of the Adler function suggest that at low and intermediate
orders the series is dominated by IR singularities. The reference model (RM) of~\cite{BJ2008}  contains 
the first two IR  and the leading UV singularities. The Borel 
transform is given by
\begin{equation}\label{RefModel}
B[\widehat D](u) \,=\, B[\widehat D_1^{\rm UV}](u)+ B[\widehat D_2^{\rm IR}](u)
+ B[\widehat D_3^{\rm IR}](u)+ d_0^{\rm PO} + d_1^{\rm PO}\,u\label{CMBJ}.
\end{equation}
The structure of the branch-cut singularities can be found in~\cite{BJ2008}. The residues and the coefficients
$ d_{0,1}^{\rm PO}$ are fixed by matching to the exactly known $c_{1,1}$ to $c_{4,1}$ (augmented by an estimate for $c_{5,1}$). 

Within this model, the conclusion of Ref.~\cite{BJ2008} in favour of FOPT has been corroborated  and 
extended in our recent work~\cite{BBJ2012}. All moments that display a good perturbative 
behaviour favour the FOPT prescription within the RM. This conclusion can be traced back
to the contribution of the leading IR singularity, related to the $D=4$ corrections in the OPE.
If this singularity is arbitrarily suppressed, one generates a model --- less realistic, in our opinion --- in which CIPT is the preferred prescription. To realize this scenario in practice, and assess possible
model dependencies in our conclusions, we introduced the following alternative model (AM) where the leading singularity  is absent whereas the sub-leading one at $u=4$ is explicitly taken into account:
\begin{equation}
\label{AltModel}
B[\widehat D](u) \,=\, B[\widehat D_1^{\rm UV}](u)+ B[\widehat D_3^{\rm IR}](u)
+B[\widehat D_4^{\rm IR}](u) + d_0^{\rm PO} + d_1^{\rm PO}\,u  \,.
\end{equation}
Within the AM, moments with good perturbative behaviour favour CIPT. 

The models represent two  quite different situations regarding  the interplay of the Adler function
coefficients and the running coupling effects.
In the RM, there are cancellations between
the contribution from the high-order coefficients $c_{n,1}$ and the running
coupling effects, at a given order in $\alpha_s$. In this case  FOPT is superior since it treats these contributions on an equal footing, while CIPT misses the cancellations due
to the resummation of the running effects to all orders.
On the other hand, the AM represents a situation where the running effects are dominant and should be 
resummed. In this case, the high-order coefficients can be neglected and CIPT is a better prescription.
Since there is no known mechanism that would naturally suppress the leading IR singularity in QCD,
we believe the  scenario of Eq,~(\ref{RefModel}) to be more realistic.

Using these two models, we compared in Ref.~\cite{BBJ2012}  the
perturbative series in FOPT and CIPT generated from 17 polynomial weight-functions $w_{i}(x)$. 
We showed that they can be divided into a small number of categories regarding 
the behaviour of their perturbative series. Generic features of the functions $w_{i}(x)$ 
(such as starting or not with the unity), together with the assumptions  upon the Adler function,
suffice to determine whether they are suitable for $\alpha_s$ extractions and whether
FOPT or CIPT is more suitable for the RG improvement. We showed that
some of the weight functions used in the literature, e.g. polynomials containing solely
powers of $x^i$ with $i\geq 2$, should be avoided due to their bad perturbative behaviour.
We also provided further arguments that support the plausibility of the RM of~\cite{BJ2008}
and concluded that for well-behaved moments FOPT is preferred.

\begin{figure}[!t]
\begin{center}
\subfigure[$w_\tau(x)=(1-x)^2(1+2x)$, FOPT.]{\includegraphics[width=.49\columnwidth,angle=0]{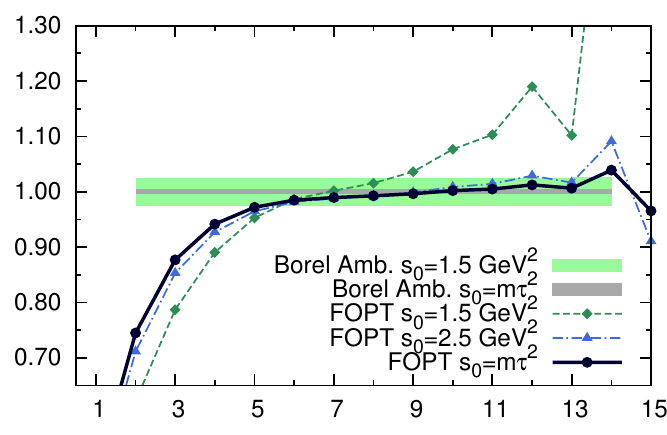}\label{RMWtauFO}}
\subfigure[$w_\tau(x)=(1-x)^2(1+2x)$, CIPT.]{\includegraphics[width=.49\columnwidth,angle=0]{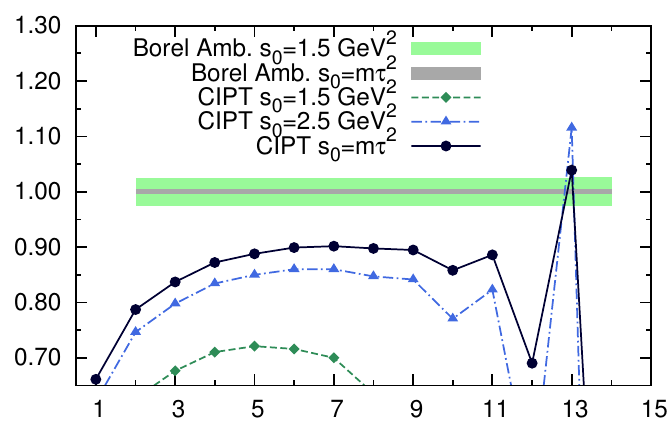}\label{RMWtauCI}}
\caption{Reference model. $\delta^{(0)}_{w_\tau}(s_0)$ order by order in $\alpha_s$ normalised to the Borel sum  for FOPT (left)  and CIPT (right) with three values of $s_0$:  1.5~GeV$^ 2$, 2.5~GeV$^ 
2$, and $m_\tau^ 2$. Bands give the Borel ambiguities. }\vspace{-0.8cm}
\label{fig:IdealMom}
\end{center}     
\end{figure}

An aspect that was not considered in~\cite{BBJ2012} was the $s_0$ dependence of these conclusions. 
This is important because sum-rules with different values of $s_0\leq m_\tau^2$ are used in  
extractions of $\alpha_s$~\cite{AlphasDVs2011,AlphasDVs2012,MY2008}. 
Here we show explicit results for the FOPT/CIPT comparison for two moments within the two models given
in Eqs.~(\ref{RefModel}) and (\ref{AltModel}) and considering three values of $s_0$: 1.5~GeV$^ 2$, 2.5~GeV$^ 
2$, and $m_\tau^ 2$. The interval [$1.5$ GeV$^2$:\ $m_\tau^2$]  spans the values used in the fits of~\cite{AlphasDVs2011,AlphasDVs2012}.
Since we intent to compare the perturbative series at different values of $s_0$ a normalisation
procedure is in order. For better comparison, we normalise the series generated for each value of $s_0$ by 
its corresponding Borel sum, Eq.~(\ref{BorelSum}). Hence, in the plots,  meaningful 
series  should be asymptotic  to the unity.

We start by considering the case of moments that have good perturbative behaviour for $s_0=m_\tau^2$. 
As a representative we choose to use the kinematic moment $w_\tau$. In 
Fig.~\ref{fig:IdealMom}, we consider the FOPT and CIPT series within the RM. On the 
left-hand side, Fig.~\ref{RMWtauFO}, one observes that the normalised FOPT series 
still behaves as a good asymptotic series even for  $s_0$ significantly smaller than 
$m_\tau^2$. As expected, for lower $s_0$, the larger values of $\alpha_s$ amplify the divergent behaviour 
 above the 8th order. Nevertheless, the first few terms of the series approach the Borel 
resummed value. Note also that the Borel sum has a larger ambiguity for smaller $s_0$ due to larger 
$\alpha_s$. On the right-hand side, in  Fig.~\ref{RMWtauCI}, one sees
that the poor performance of CIPT is amplified by the larger values of the coupling at lower $s_0$. 
That is, within the RM, CIPT is not a good approximation to the Borel resummed values, and even less
so for smaller $s_0$.

\begin{figure}[!t]
\begin{center}
\subfigure[$w_{17}(x)=(1-x)^2x^3(1+2x)$, FOPT.]{\includegraphics[width=.49\columnwidth,angle=0]{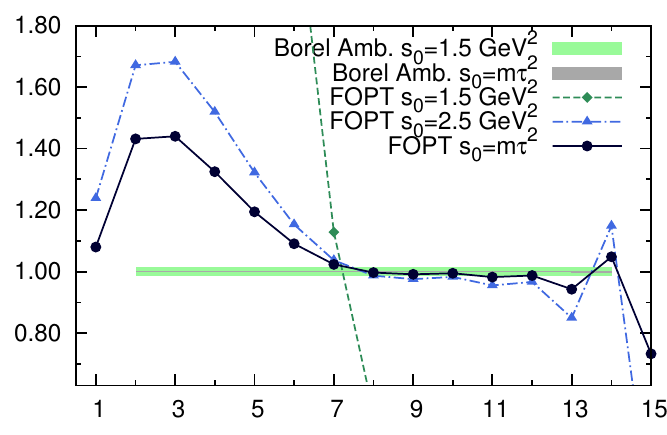}\label{RMW17FO}}
\subfigure[$w_{17}(x)=(1-x)^2x^3(1+2x)$, CIPT.]{\includegraphics[width=.49\columnwidth,angle=0]{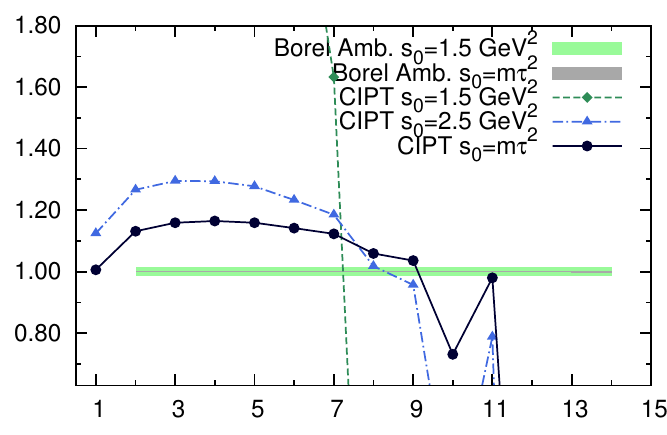}\label{RMW17CI}}
\caption{Reference model. $\delta^{(0)}_{w_{17}}(s_0)$ order by order in $\alpha_s$ normalised to the Borel sum  for FOPT (left)  and CIPT (right) with three values of $s_0$:  1.5~GeV$^ 2$, 2.5~GeV$^ 
2$, and $m_\tau^ 2$. Bands give the Borel ambiguities.}\vspace{-0.7cm}
\label{fig:ProMoments}
\end{center}     
\end{figure}
We now turn to a moment with bad perturbative behaviour: $w_{17}(x)=(1-x)^2x^3(1+2x)$ (to employ the 
notation of~\cite{BBJ2012}). In 
Ref.~\cite{BBJ2012} we showed that moments starting with powers of $x$ (that do not contain the unity)
tend to have bad perturbative behaviour and are largely dominated by power corrections. In 
Fig.~\ref{fig:ProMoments} we address the $s_0$ dependence of this conclusion. On the left, 
Fig.~\ref{RMW17FO} shows that for higher values of $s_0$ FOPT can approach the Borel result only at high 
orders (not available exactly). At low $s_0$ the series displays a wild behaviour and cannot 
be considered a good approximation to the Borel sum. In CIPT, Fig.~\ref{RMW17CI}, the bad behaviour 
already observed for $s_0=m_\tau^2$ is amplified at lower $s_0$. The series are erratic and cannot
be consider suitable asymptotic approximations to the Borel sum. Note that this moment, despite of its bad perturbative behaviour, enters several  determinations of $\alpha_s$ from $\tau$ decays (e.g. Refs.~\cite{ALEPH,OPAL}).
\begin{figure}[!ht]
\begin{center}
\subfigure[$w_\tau(x)=(1-x)^2(1+2x)$, FOPT.]{\includegraphics[width=.49\columnwidth,angle=0]{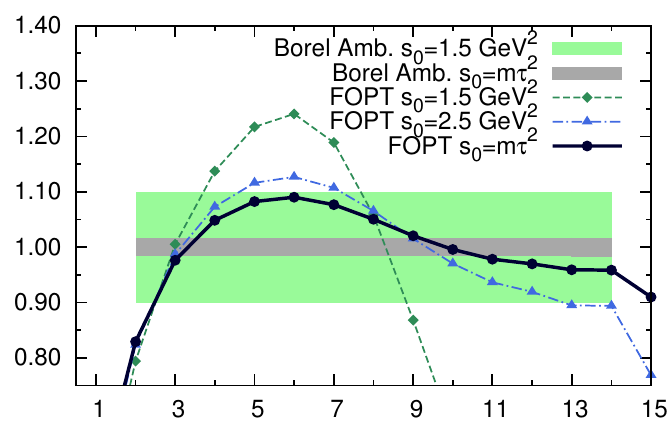}\label{AMWtauFO}}
\subfigure[$w_\tau(x)=(1-x)^2(1+2x)$, CIPT.]{\includegraphics[width=.49\columnwidth,angle=0]{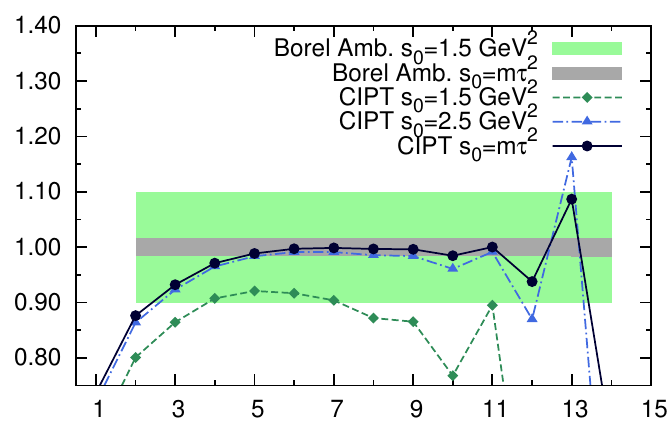}\label{AMWtauCI}}
\caption{Alternative model. $\delta^{(0)}_{w_{\tau}}(s_0)$ order by order in $\alpha_s$ normalised to the Borel sum  for FOPT (left)  and CIPT (right) with three values of $s_0$:  1.5~GeV$^ 2$, 2.5~GeV$^ 
2$, and $m_\tau^ 2$. Bands give the Borel ambiguities.}
\label{fig:IdealMomAM}
\end{center}     
\end{figure}

We can perform the same analysis in the alternative model, Eq.~(\ref{AltModel}), which receives no
contribution from the leading IR singularity. In Fig.~\ref{fig:IdealMomAM}, we
show the series normalised to their respective Borel resummed values within the AM for $w_\tau(x)$.
In this model, CIPT provides the better framework also for lower values of $s_0$, as shown in 
Fig.~\ref{AMWtauCI}. The series remains very stable for $s_0=2.5$ GeV$^2$ and still approaches the Borel result well. For $s_0=1.5$ GeV$^2$ CIPT can still be considered a good approximation taking into account the amplified Borel ambiguity. 
The oscillations of FOPT, already present at $s_0=m_\tau^2$, are much amplified for lower $s_0$ (see 
Fig.~\ref{AMWtauFO}). Within the AM, the FOPT series are not a good approximation to the Borel resummed values.

\begin{figure}[!ht]
\begin{center}
\subfigure[$w_{17}(x)=(1-x)^2x^3(1+2x)$, FOPT.]{\includegraphics[width=.49\columnwidth,angle=0]{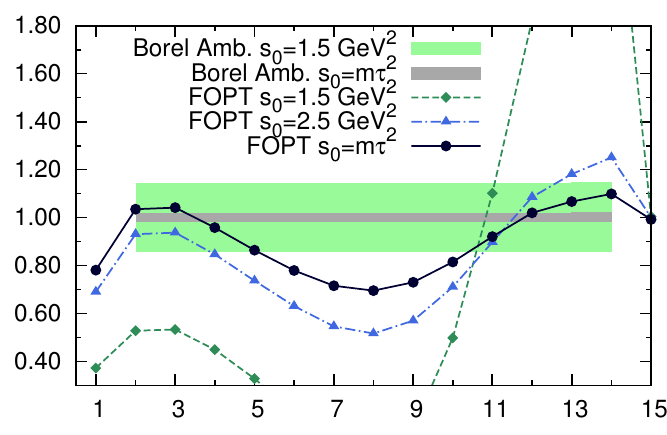}}
\subfigure[$w_{17}(x)=(1-x)^2x^3(1+2x)$, CIPT.]{\includegraphics[width=.49\columnwidth,angle=0]{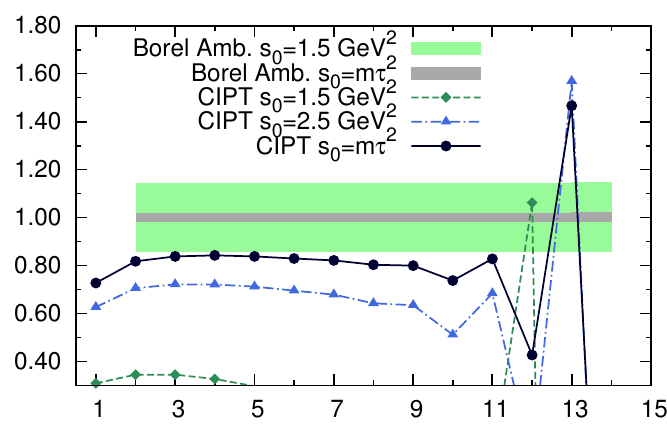}}
\caption{Alternative model. $\delta^{(0)}_{w_{17}}(s_0)$ order by order in $\alpha_s$ normalised to the Borel sum  for FOPT (left)  and CIPT (right) with three values of $s_0$:  1.5~GeV$^ 2$, 2.5~GeV$^ 
2$, and $m_\tau^ 2$. Bands give the Borel ambiguities.}\vspace{-0.3cm}
\label{fig:ProMomentsAM}
\end{center}     
\end{figure}

To conclude we examine the case of $w_{17}$ in the context of the AM. The results are shown in
Fig.~\ref{fig:ProMomentsAM}. The bad perturbative behaviour of FOPT and CIPT remains
for all values of $s_0$. This is an indication of the model independence of the conclusion 
that $w_{17}$ (and a number of other moments also discussed in Ref.~\cite{BBJ2012}) should be avoided in determinations of $\alpha_s$. 
 
\section{Conclusions}

Recently, we have analysed the perturbative behaviour of several moments 
often employed in analyses of $\alpha_s$ from $\tau$ decays under different
assumptions for the large-order behaviour of the Adler function~\cite{BBJ2012}.
We have shown that some of these moments should be avoided 
due to their bad perturbative behaviour. Furthermore, under reasonable assumptions
for the Borel transformed Adler function, we showed that FOPT provides
the preferred framework for the RG improvement of moments that display good perturbative
behaviour. 

Here we showed, for the first time, that these conclusions are still valid
if one considers the perturbative series generated by FOPT and CIPT for $s_0\leq m_\tau^2$.
This is a relevant question, since in $\alpha_s$ extractions one often considers sum-rules with $s_0\leq m_\tau^2$. We have shown
explicitly the results for two representative  moments previously investigated in Ref.~\cite{BBJ2012}
for $s_0=m_\tau^2$. The $s_0$ dependence analysis was also carried  out for the remaining moments 
studied in~\cite{BBJ2012} with similar conclusions; they are not shown here for the sake of
brevity.

\section*{Acknowledgements}
The author wishes to thank Martin Beneke and Matthias Jamin for the careful reading of the manuscript. This work was supported by the Alexander von Humboldt Foundation.

\end{document}